# Reliability-Aware Overlay Architectures for FPGAs: Features and Design Challenges

Mihalis Psarakis
Dept. of Informatics
University of Piraeus, Greece
mpsarak@unipi.gr

*Abstract*— **The FPGA overlay architectures have been mainly proposed to improve design productivity, circuit portability and system debugging. In this paper, we address the use of overlay architectures for building fault tolerant SRAM-based FPGA systems and discuss the main features and design challenges of a reliability-aware overlay architecture.**

## I. INTRODUCTION

FPGAs have gained an increasing acceptance the last decades for building embedded systems for critical applications, such as avionics, space, automotive, medical, etc. where the high performance requirements are combined with strict reliability standards. However, due to their aggressive circuit scaling, SRAM-based FPGAs have become extremely vulnerable to various failure mechanisms causing transient and permanent errors, with the radiation-induced single-event upsets (SEUs) being the major cause of failure. Although several efficient fault tolerance approaches for SRAM-based FPGAs have been proposed in the past [1], the design of reliable FPGA-based systems for critical and harsh-environment applications is still an active research topic [2].

The adoption of overlay architectures in the design flow of reliable FPGA-based systems will provide the typical benefits presented so far in the literature for the use of FPGA overlays, i.e. improved productivity, circuit portability, debugging, etc., but it will also help FPGA designers to apply efficient fault tolerance techniques. For example, when an FPGA overlay is employed to build a fault tolerant design using the triple-modular redundancy (TMR) technique, the FPGA designer does not need to know in depth its implementation details, since they would have been integrated in the design features of the specific overlay architecture. In this paper, we investigate the use of *reliability-aware overlay architectures* for building SRAM-based FPGA designs. First, we analyze the benefits of using a customized FPGA overlay architecture in order to apply typical fault tolerance techniques and then we present the main features and the design challenges of a reliability-aware overlay architecture.

Let's assume that we have decided to use an overlay-based design flow to build an FPGA system for a critical application and we plan to improve the system reliability applying the TMR technique. There are two design compilation options: (a) to employ an existing TMR tool, e.g. Xilinx TMRtool, during synthesis step in order to generate a triplicated RTL design and then rely on the original overlay architecture and the corresponding design flow and (b) to modify the original overlay architecture in order to integrate TMR-related features and adapt the corresponding tool chain (overlay mapping, placement and routing algorithms) in order to directly map the original RTL design into the modified overlay fabric. The advantages of solution (b) against solution (a) are: (i) less hardware overhead and performance degradation due to the integration of fixed logic (e.g. majority voters) in the modified overlay architecture and less interconnection resources and (ii) improved reliability due to the integration of placement rules during the overlay fabric implementation to avoid phenomena, such as single SEUs affecting more than one triplicated modules which may cancel the fault masking property of the TMR technique.

## II. RELIABILITY-AWARE OVERLAY ARCHITECTURE

In this Section, we present the main features and the design challenges of a reliability-aware overlay architecture. Notice that we do not propose a new overlay architecture but the modification of an existing one. In order to demonstrate the proposed concept, we adopt the FPGA overlay architecture proposed by Stitt and Coole in [3], called *intermediate fabric (IF)*. IF is a coarse-grain overlay architecture specialized for image processing kernels. Notice that the proposed reliability-aware overlay concept can be potentially integrated in most existing overlay architectures.

### A. Reliability-aware Intermediate Fabric

We focus on the datapath component of the IF architecture which contains arithmetic resources (multipliers, subtractors with an optional absolute-value function, adders, etc.) commonly used in image-processing kernels, such as 2D convolution (shown in Fig.1a), sum-of-absolute differences (SAD), Sobel edge detectors, etc. An island-type datapath fabric for the IF architecture [3] consisting of functional units (FUs), switch boxes and connection boxes is shown in Fig.1b.

Let's assume that we want to apply the TMR technique at RT-level to the 2D convolution kernel, as shown in Fig.2a. A naïve solution would be to integrate the required number of voters to the datapath architecture of the original IF. In this paper, we propose the modification of the FUs of the datapath architecture as shown in Fig.2b in order to be triplicated and contain a dedicated delay-optimized majority voter.

The publication of this paper has been partly supported by the University of Piraeus Research Center.

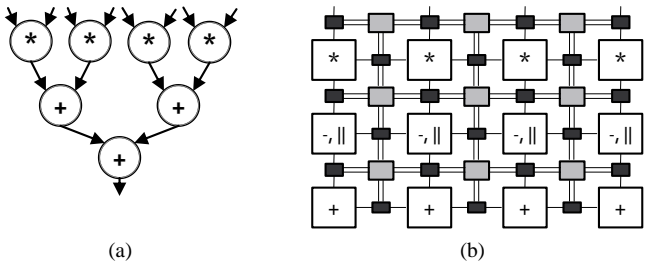

Fig. 1. (a) 2x2 2D convolution kernel, (b) Island-type datapath architecture.

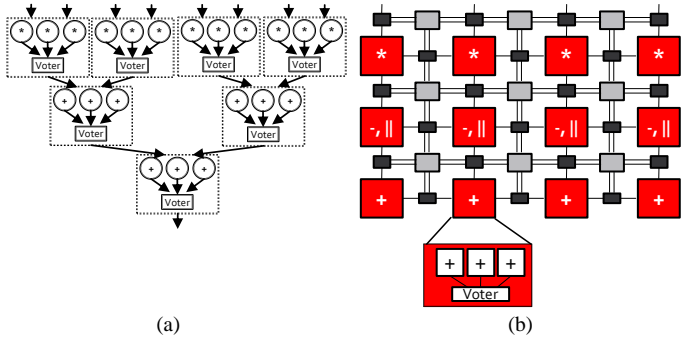

Fig. 2. (a) TMR 2D convolution kernel, (b) Reliability-aware datapath.

To demonstrate the efficiency of the proposed approach in terms of hardware overhead, we implemented two different overlay architectures that fit the design of 2x2 2D convolution and SAD kernels: (a) the naïve approach, which requires 12 multipliers, 12 subtractors/abs, 9 adders and 7 voters, as shown in Fig.2a, and (b) the proposed approach, which requires 4 TMR multipliers, 4 TMR subtractors/abs and 3 TMR adders, as shown in Fig.2b. The area utilization of the two overlays for a Virtex-5 XC5VLX110T device is shown in Table I. The significant lower area overhead of the proposed approach is mainly due to the smaller array - and thus the smaller number of connection and switch boxes which occupy a large portion of the overlay fabric area - compared to the array of the naïve approach. It has to be noted that the area utilization results are preliminary since the overlay designs are non-optimized. Furthermore, the current version of the proposed architecture protects only the functional units through the TMR technique and not the connection and switch boxes. In the future, we plan to present an optimized reliability-aware overlay architecture including the hardening of the routing resources.

TABLE I. OVERLAY FABRIC AREA UTILIZATION

| Approach | LUTs | FFs | DSPs |
|---|---|---|---|
| Naïve | 32% | 12% | 18% |
| Proposed | 14% | 5% | 18% |

### B. Design Challenges of a Redundant Overlay Fabric

Besides the triplication of the FUs and the integration of majority voters, the redundant overlay fabric should comprise the following features:

- The design of the triplicated FUs must take into consideration the design rules of previously proposed TMR schemes [4] and voters [5] to guarantee the protection of FPGA logic, registers, routing, memories, state machines, etc. against soft errors.

- The mux-based connection and switch boxes must be efficiently protected against soft errors in the FPGA configuration bits without using triplication, since the adoption of TRM technique for these modules would explode the area overhead.

- More redundancy-based approaches must be examined for the protection of the functional units, e.g. duplication-with-comparison (DWC), error detection and correction codes (EDCs). The overlay fabric should contain different types of hardened FUs, in order to facilitate the building of area-efficient mixed critical systems. For example, in a mixed critical system, TMR FUs could be used for the most critical modules while DWC FUs may be adequate for the less critical modules. Furthermore, the FUs with EDCs could be used to enable the hardening of arithmetic applications at algorithmic level.

- The redundant modules of the overlay fabric must be physically separated to avoid the phenomenon of single SEUs affecting more than one replicas. Placement constraint rules could be adopted from existing approaches [6].

### C. Two-level Scrubbing Approach

The scrubbing technique is used to detect and correct soft errors in the FPGA configuration memory. The scrubbing process for an overlay-based FPGA design should be applied at two levels: (a) scrubbing of the FPGA configuration memory (lower level) and (b) scrubbing of the overlay configuration memory (upper level).

At the lower level, the scrubbing process depends on the FPGA device architecture and uses the EDC codes embedded in the FPGA configuration frames. Several approaches have been recently presented to speedup the FPGA configuration memory scrubbing process [7], [8] and can be adopted in our case. Since the design placement may affect the execution time of the scrubbing process [7], the placement of the overlay fabric should be considered for speeding up the scrubbing.

At the upper level, the scrubbing process is device independent and has shorter execution time compared than the lower scrubbing level. To support runtime error detection and correction at upper level, we must embed EDC codes in the configuration memory of the overlay architecture. The scrubbing process at both levels can be efficiently driven by an accurate soft error sensitivity analysis of the FPGA system, which is discussed in the next subsection. Another issue is the design of a hardened controller for the fast and reliable scrubbing of the overlay configuration memory.

### D. Soft Error Sensitivity Analysis

Soft error sensitivity analysis is mainly used to classify the FPGA configuration bits into sensitive and non-sensitive depending on the impact of the corresponding soft errors to the circuit behavior. In our case, the sensitivity analysis must be performed at two levels, similarly to the scrubbing process: (a) at the FPGA configuration memory and (b) at the overlay configuration memory. At the lower level, the sensitivity analysis depends on both the FPGA and the overlay architecture. It must be performed once for the FPGA-overlay pair and can be fine-tuned after the implementation of the specific FPGA design taking into consideration the utilized overlay resources. The existing fault injection platforms can be used for the lower level sensitivity analysis. On the contrary, at the upper level, the sensitivity analysis is independent of the FPGA architecture and requires the devise of new methodologies and tools.

*E. Spare Resources for Active Fault Repair*

To support active fault repair, we could consider the allocation of spare resources in the mapping of FUs and interconnection network of the reliability-aware overlay architecture. These spare resources can be used to restructure the overlay fabric by reconfiguring the FPGA in order to adapt to the emergence of permanent faults. In the FPGA literature, such methods are separated into: (a) a-priori resource allocation approaches, where spare resources are assigned during design time and alternative precompiled configurations are generated each one using a different subset of programmable resources, and (b) dynamic approaches which diagnose the faulty resources and devise the alternative FPGA configuration at runtime. In our case, precompiled configurations can be generated for the programmable resources of the overlay fabric. The precompiled configurations can be used to reconfigure either the entire overlay architecture or part of it, e.g. a functional unit and its replicated modules. Various parameters must be taken into consideration, such as the portion of spare resources, the reconfiguration granularity, the number of precompiled configurations to achieve the optimal compromise between repair time, area overhead and fault repair capacity.

In case of conventional FPGA systems, due to the lack of low-level on-line synthesis tools and the fact that the FPGA bitstream cannot be openly generated, the implementation of dynamic approaches is not feasible for the current FPGA technology. However, the use of overlay architectures will enable the development of lite, online implementation tools and thus the development of dynamic fault repair approaches.

III. CONCLUSION AND FUTURE WORK

We presented a reliability-aware overlay architecture for SRAM-based FPGAs and demonstrated it to a coarse-grain overlay architecture for image-processing applications. We also discussed several design challenges and research issues raised by such a prospect. In the future, we plan to optimize the proposed overlay architecture, protect the remaining modules (e.g. connection and switch boxes) and implement the corresponding tool chain. To evaluate the efficiency of the proposed approach, we aim to compare it against the following alternative options for a set of image-processing kernels in terms of area overhead, performance degradation and compilation time: (a) TMR kernel generated by a commercial TMR tool and implemented without using overlay and (b) RT-level TMR kernel implemented using the original overlay architecture enhanced with voters (naïve approach).